\documentclass[aps,reprint]{revtex4-1}
\usepackage{graphicx}
\usepackage{mathtools}
\usepackage{bm}
\usepackage{amssymb}   
\usepackage{color} 

\draft 

\begin{document}


\title{Zealotry Effects on Opinion Dynamics in the Adaptive Voter Model}


\author{Pascal P. Klamser}
\affiliation{Potsdam Institute for Climate Impact Research, P.O. Box 60 12 03, 14412 Potsdam, Germany, EU}
\affiliation{Department of Physics, Humboldt University, Newtonstr. 15, 12489 Berlin, Germany, EU}
\author{Marc Wiedermann}
\affiliation{Potsdam Institute for Climate Impact Research, P.O. Box 60 12 03, 14412 Potsdam, Germany, EU}
\affiliation{Department of Physics, Humboldt University, Newtonstr. 15, 12489 Berlin, Germany, EU}
\author{Jonathan F. Donges}
\affiliation{Potsdam Institute for Climate Impact Research, P.O. Box 60 12 03, 14412 Potsdam, Germany, EU}
\affiliation{Stockholm Resilience Centre, Stockholm University, Kr\"aftriket 2B, 114 19 Stockholm, Sweden, EU}
\author{Reik V. Donner}
\affiliation{Potsdam Institute for Climate Impact Research, P.O. Box 60 12 03, 14412 Potsdam, Germany, EU}


\date{\today}

\begin{abstract}
The adaptive voter model has been widely studied as a conceptual model for opinion formation processes on time-evolving social networks.
Past studies on the effect of zealots, i.e., nodes aiming to spread their fixed opinion throughout the system, only considered the voter model on a static network. 
Here, we extend the study of zealotry to the case of an adaptive network topology co-evolving with the state of the nodes and
investigate opinion spreading induced by zealots depending on their
initial density and connectedness. Numerical simulations reveal that
below the fragmentation threshold a low density of zealots is sufficient to
spread their opinion to the whole network. Beyond the transition point,
zealots must exhibit an increased degree as compared to ordinary nodes 
for an efficient spreading of their opinion.
We verify the numerical findings using a mean-field approximation of the model yielding a 
low-dimensional set of coupled ordinary differential equations. Our results imply that the spreading of the zealots' opinion in the adaptive
voter model is strongly dependent on the link rewiring probability and the average
degree of normal nodes in comparison with that of the zealots. In order
to avoid a complete dominance of the zealots' opinion, there are two possible strategies for the remaining nodes: adjusting the probability of rewiring and/or the number of connections with other nodes, respectively.
\end{abstract}


\maketitle 

\section{Introduction}

The study of opinion dynamics on social networks is a popular application of network 
and complex systems theory~\cite{Albert2002, Newman2003}.
Among others, the voter model (VM) is a widely investigated idealized model describing the dynamical behavior of individual opinions on social networks and
represents a bridge between instructive toy models in physics and social science~\cite{Mobilia2003, Mobilia2007, Amblard2004, Franks2008}.
Closely related to the VM are epidemic models on a network structure, where the health state of an individual (node) takes the role of a discrete opinion~\cite{Selley2014, Gross2006, Pastor-Satorras2001, May2001}.


The problem of driving a given system to a desired state (like a certain consensus opinion) is commonly addressed by concepts of control theory~\cite{Liu2011a, Yuan2013, Cowan2012, Nacher2012}. Specifically, the problem of network controllability was reformulated as an eigenvalue problem \cite{Yuan2013}, the applicability of which to real-world networks has been discussed in \cite{Cowan2012}. As a result, a simple control strategy for heterogeneous networks was proposed~\cite{Nacher2012}. From this problem setting, interesting concepts arose (such as the maximum matching set, i.e., the maximum set of links which do not share the same start and end nodes) to identify the minimum set of driver nodes to structurally control the whole network~\cite{Liu2011a}. Surprisingly, with this procedure hub nodes are avoided as driver nodes.

From the perspective of opinion dynamics and, more specifically, the VM, one way to conveniently study the problem of controlling (social) network dynamics is by introducing zealots to the system. Here, zealots are stubborn agents who either favor \cite{Mobilia2003} or fully maintain \cite{Mobilia2007} one specific opinion. Both types of zealotry have been extensively studied in the context of different opinion formation models \cite{Mobilia2003,Mobilia2007,Galam2007,Yildiz2013}. In the present study, we take the latter viewpoint and define zealots as nodes of a network that never change their dynamical state (opinion) during the evolution of the VM. In analogy to control theory, this type of zealots can be seen as nodes receiving an external input signal which pins their state~\cite{Sorrentino2007a}. It might be interesting to note that extremists in bounded confidence opinion models can also be seen as a weak form of zealots~\cite{Amblard2004, Franks2008}. In contrast to fully stubborn zealots as studied in the present work, they can still change their opinion, but this process is very unlikely as compared to other nodes.

In the context of the classical VM, the effect of zealotry has already been studied on regular lattices with a single zealot~\cite{Mobilia2003} and with a finite number of zealots on regular, complete~\cite{Mobilia2007} and random graphs~\cite{Xie2011}. The latter study found a transition at a specific density of zealots where the time to reach consensus was drastically decreased. Moreover, the optimal topological placement of zealots was investigated~\cite{Yildiz2013, Wu2004}, whereby high-degree hubs were found to be good positions from where to spread the opinion. Similar observations have also been made in bounded confidence models with extremists~\cite{Amblard2004, Franks2008}, in which hubs were found to be good placements for extremists to bias the overall opinion. Moreover, other opinion formation models like the majority voter model \cite{Galam2007} are also known to exhibit particularly rich dynamics on introduction of zealots.

All aforementioned studies assumed dynamics on a fixed network structure. However, the results obtained do not apply to systems with a time-evolving network topology ~\cite{Holme2015}.
Adaptive network models like the adaptive voter model (AVM)~\cite{Holme2006} or other more realistic models of opinion formation~\cite{Schleussner2016a} generate a time-dependent network structure through a feedback mechanism between topology and node (agent) states. The AVM extended the classical VM by giving the nodes the possibility to break an existing link and reconnect to a like-minded node which results in a temporally evolving network. The standard AVM has been analyzed in terms of mean-field theory which revealed the existence of two absorbing states, namely the active and frozen (fragmented) state, where for finite systems, the active state asymptotically becomes the consensus state (exhibiting a giant component with a single opinion only) \cite{Vazquez2008b,Kimura2008}. Further, it was found that minor changes in the microscopic update rules can reduce or increase the time necessary to reach a final state due to network topology-state feedbacks \cite{Nardini2008}. 

In the context of temporal networks (i.e., networks with links that are only present intermittently, including the adaptive networks investigated in the present work as a specific class), controllability was investigated, alongside other studies, by a time-respecting path-based method~\cite{Posfai2014} and by an analytical approach combined with graphical tools~\cite{Pan2014}. The latter study revealed a positive relation between the aggregated degree of a node, the number of interactions during a given time, and the size of the subset which is controlled by it. Both studies quantified the controllable subset by the influence of a single node, assuming linear dynamics and considering networks which are statistically equivalent at different times. However, it has remained unclear so far how the control of the AVM could be best achieved or avoided, because its dynamics are strongly nonlinear and the network is in certain parameter regimes evolving in such a way that it is not statistically equivalent at different times. Furthermore, previous studies on network controllability commonly addressed only temporal networks without feedback between topology and node state, which is a key property of the AVM. Despite the resulting differences between the AVM and controllability studies on other temporal networks, we are confident that the combination of both aspects is a very promising field of research. To our best knowledge, the concept of zealots, widely studied in the static VM, has not yet been applied in the AVM. Since the latter is a relevant conceptual model for opinion formation on temporal networks, which obeys relatively simple rules allowing for a fair degree of understanding of the resulting dynamics, we see a great interest in addressing the issues mentioned above.

Consequently, this paper addresses the efficiency of control by zealot opinion spreading (ZOS) in an extension of the AVM. Here, the zealots are chosen at random and possess additional links and therefore an excess degree compared to ordinary nodes, which is beneficial for the spreading, as former related studies suggest~\cite{Yildiz2013, Wu2004, Amblard2004, Franks2008, Pan2014}. The excess degree is not only motivated by the expected effect on the spreading but also from real world examples~\cite{Enos2016}. For instance, election campaigns aim to reach as many voters (nodes) as possible. Following their mission, campaigners are not convinced by voters, there is only a unidirectional influence of campaigners on voters. Additionally, campaigners reach an effectively increased degree in social networks due to their professional outreach efforts~\cite{Enos2016}. Another example for zealotry are lobbyists intervening in political processes.
The number of zealots and their excess degree can be interpreted as a measure of the resources that have been invested to pursue the campaign. Another important issue is how easily a system can be controlled and what needs to be changed in order to increase its resilience against external pressure or corruption. 

After the description of the model and methods in Sect.~\ref{sec:pre}, we focus on the question how the zealot opinion is spread over clusters of different sizes in Sect.~\ref{sec:cluster}. Thereby, we observe the emergence of subgraphs with a significantly larger mean degree than that of the whole network. We identify three different parameter regimes, closely related to the phase transition in the AVM, in which different effects lead to a significant increase in the spreading efficiency of the zealots' opinion. Our numerical microscopic results are further supported by an analytical macroscopic approximation of the AVM including zealots (Sect.~\ref{sec:ana}). Ultimately, we explore the consequences of two different adaptation rules, in which the zealots either do not rewire at all or obey heterophilic rewiring only (Sect.~\ref{sec:altup}). All obtained results are discussed and conclusions drawn in Sect.~\ref{sec:discuss}.

\section{Methods \label{sec:pre}}

\subsection{Model description \label{sec:model}}

We study the AVM in the version originally formulated by Holme and Newman~\cite{Holme2006}, but thoroughly extended by introducing zealots with excess degree. Here, the two processes governing the opinion dynamics in a network with $N$ nodes, also referred to as voters, and $G$ different opinions are the change of node opinions and the link rewiring process, the mathematical formulations of which will be presented in the following. As the most crucial parameter of the resulting adaptive network model, the rewiring probability $\phi$ is considered as the fraction of cases in which the latter process takes place instead of the former. Each node $i$ initially possesses an opinion $g_i$, which is on average the opinion of a number of $\gamma_0=N/G$ nodes. For each node $i$, we define the set of nodes $\mathcal{S}_i= \mathcal{S}_{g_i} \backslash \{ \{i\} \cup \mathcal{N}_i\}$ by excluding from the set $\mathcal{S}_{g_i}$ of nodes having opinion $g_i$ the node $i$ itself and the set of its direct neighbors $\mathcal{N}_i=\{j: A_{ij}=1\}$ (here, $A_{ij}$ denote the entries of the network's adjacency matrix at a given point in time, where we have suppressed the associated time index for brevity). 

The dynamic update cycle of the model is then described as follows:\\

\begin{minipage}{0.95\columnwidth}
\noindent
Step 1: Randomly select a node $i$. If the degree $k_i$ of node $i$ is zero do nothing, otherwise randomly select a neighbor $j\in\mathcal{N}_i$. \\
Step 2(a): With probability $\phi$, delete the link to $j$ and rewire to a randomly selected node $j^*\in\mathcal{S}_i$ with $j^*\neq j$ (rewiring). If $|\mathcal{S}_i|=0$ do nothing.\\
Step 2(b): With probability $(1-\phi)$, node $i$ imitates the opinion of node $j$ and, thus, $g_i\to g_j$ (imitation). \\
\end{minipage}

In contrast to the classical formulation of the AVM, in this formulation no multiple links and self-loops are possible due to node rewiring to the set $\mathcal{S}_i$ (instead of rewiring to the set $\mathcal{S}_{g_i}$ as in the standard AVM).
Note that step 2 is applied regardless of an existing opinion conflict.
The algorithm is iteratively repeated until a time $t_c$, where the final state is reached in which only like-minded nodes are connected to each other ($\mathcal{N}_i\subseteq \mathcal{S}_{g_i}\ \forall i$).

It shall be stressed, that the rules employed in this model variant are based on node selection~\cite{Holme2006, Kimura2008, Wiedermann2015} instead of link selection~\cite{Silk2014, Bohme2011, Demirel2014}, because we consider it more realistic 
for a social network that agents (nodes) spend on average the same time for communicating with others. The number of nodes $N$ and the total number of links $M$ stay constant over time, which implies that the mean degree $\overline{k}_0=\frac{2M}{N}$ of the network is kept fixed.

%
%
As already mentioned in the introduction, in this work, we consider zealots as nodes that cannot be convinced. For the sake of simplicity, we assume here that all zealots carry the same specific opinion $g_z$. The set of zealots $\mathcal{S}_z$ is created at the start of each simulation by randomly declaring a fraction of nodes $n_z^0$ to be zealots. Their key property of having an immutable opinion is ensured by modifying the last step in the above scheme:

\vspace{0.2cm}
\begin{minipage}{0.95\columnwidth}
\noindent
Step 2 (b) With probability $(1-\phi)$, if $i\in \mathcal{S}_z$ do nothing, otherwise node $i$ adopts the opinion of node $j$.
\end{minipage}
\vspace{0.2cm}

In comparison with all non-zealot nodes, the initial mean degree of zealots can be further increased by an excess degree $k_x$ to $\overline{k}_z(t=0)=\overline{k}_0+k_x$, which describes additional links that are randomly connecting each zealot to non-zealots. This excess degree (which is considered here to be the same for each zealot to simplify the following analyses) is motivated by the additional efforts of campaigners or lobbyists to convince as many nodes as possible. Starting with a specific configuration, the degree of individual zealots then changes over time according to the considered rules of the update cycle. Let the initial fraction of nodes holding opinion $g_z$ be $n_{g_z}(0)=n_z^0$. To simplify the notation, $n_{g_z}$ will be denoted in the following as $n_z=n_{g_z}$. Note that introducing the excess degree $k_x$ increases the mean degree of the whole network to $\overline{k}=\overline{k}_0+2n_z^0k_x$, and the introduction of zealots changes the average number of nodes initially holding a certain opinion different from $g_z$ to $\gamma = \gamma_0(1-n_z^0)$. 

\subsection{Perspectives on zealot opinion spreading \label{sec:spread}}

Zealots are nodes with excess degree intending to spread their fixed opinion $g_z$ to as many other non-zealot nodes as possible. In this special case, we investigate the zealot opinion spreading (ZOS) process characterized by the fraction of nodes $n_z(t)$ holding the zealots' opinion, which is a special opinion since it is (unlike the others) always present due to the zealots that cannot become convinced by others.
The ZOS efficiency is defined as the fraction of nodes $n_z(t_c)$ holding opinion $g_z$ when the final state is reached.

As emphasized above, the considered problem can also be viewed from a controllability perspective. Here, zealots are just normal nodes that are influenced by a constant control signal $b(t)=b$ which fixes their opinion to $g_z$. The excess degree $k_x$ of zealots can be viewed as a ``topological input signal'' applied to the network only once at $t=0$. In real world applications, constraints exist in terms of resource limitations (campaigners or lobbyists need to be paid) or ideology (not everyone wants to be a zealot), which motivates us to limit the number of nodes which can receive the said control signal $b(t)$.

%
%
%
\begin{figure}
    \centering
    \includegraphics[width=1.0\linewidth]{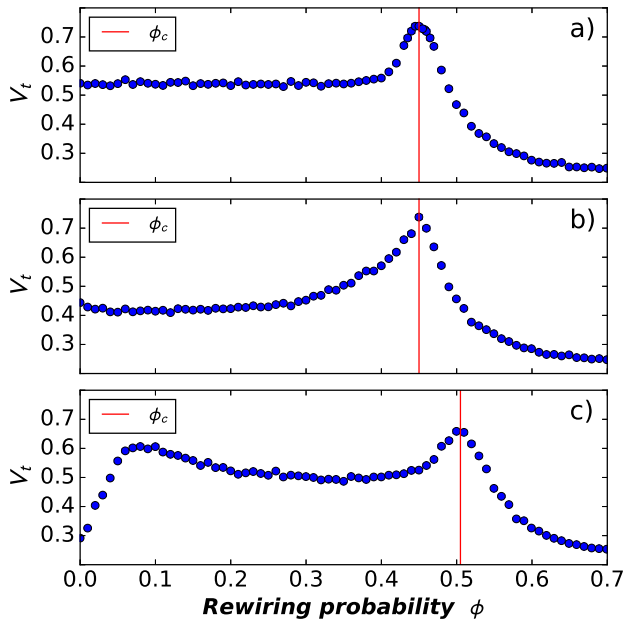}
\caption{(Color online) Coefficient of variation of convergence time $V_t$, indicating a phase transition at $\phi=\phi_c$ where the giant component vanishes. The different parameter settings are (a) no zealots $n_z^0=0$, (b) zealot density $n_z^0=0.01$ with no excess degree $k_x=0$, and (c) zealot density $n_z^0=0.01$ with an excess degree of $k_x=20$. The red lines in (a,b) at $\phi_c=0.45$ and in (c) at $\phi_c=0.505$ indicate the approximate positions of the phase transition, which is shifted if the excess degree $k_x$ is nonzero. Distributions are computed from $n=10^4$ runs on graphs with $N=800$ nodes and a mean degree (without the zealots' excess degree) of $\overline{k}_0=4$.}
    \label{fig:1}
\end{figure}
%

\section{Numerical results \label{sec:cluster}}

The AVM, obeying the discrete-time update rules as described in Section~\ref{sec:model}, starts evolving at $t=0$ from an Erd\H{o}s-R\'enyi random graph, which is known to exhibit a giant component if $\overline{k}_0\ge 1$~\cite{Erdos1959, Albert2002}. A phase transition occurs when increasing the rewiring probability $\phi$ to a level at which the giant component of the initial network vanishes and homogeneous clusters emerge which are formed by like-minded nodes. This fragmentation transition occurs at a critical rewiring probability $\phi_c$ (for fixed mean degree $\overline{k}$) or at a critical mean degree $\overline{k}_c$ (for fixed $\phi$). In what follows, we will mostly follow the strategy of varying $\phi$ but keeping the mean degree fixed.

\subsection{Fragmentation transition}

One way of identifying the fragmentation transition and the associated parameter value $\phi_c$ is based on critical slowing down~\cite{Holme2006} indicated here by a maximum of the coefficient of variation $V_t=\sigma_{t_c}/\overline{t}_c$ of the convergence time as a characteristic parameter, where $\overline{t}_c$ and $\sigma_{t_c}$ denote the empirical mean value and standard deviation of the convergence time estimated from a sufficiently large ensemble of independent realizations of the considered AVM variant~\cite{Holme2006}. In the present work, $V_t$ is estimated as a function of $\phi$ for $N=800$, $\overline{k}_0=4$, $\gamma_0=10$ and $10^4$ simulation runs (Fig.~\ref{fig:1}). If no zealots exist (Fig.~\ref{fig:1}(a)), the phase transition occurs at $\phi_c\approx 0.45$, as expected from previous studies~\cite{Holme2006}. If only few zealots are introduced ($n_z^0=0.01$) (Fig.~\ref{fig:1}(b)) the phase transition point is not altered. However, declaring them as hubs with an additional excess degree of $k_x=20$ (Fig.~\ref{fig:1}(c)) shifts the phase transition to $\phi_c=0.505$. Since the mean degree $\overline{k}$ has a strong impact on the transition~\cite{Holme2006}, this shift in $\phi_c$ can be explained by its change $\overline{k}_0\to\overline{k}=4.4$. Simulations with $\overline{k}_0=4.4$
and no zealots (not shown) indeed perfectly reproduce the observed shift, implying the redistribution of the additional links to the whole network in $t\le t_c$.

Another phenomenon is observed if the zealots' excess degree is increased, which manifests in a second broad maximum of $V_t$ in Fig.~\ref{fig:1}(c) at about $\phi=0.08$. This secondary maximum is shifted towards larger $\phi$ if $k_x$ is further increased (not shown) and is explained by the assortativity in degree, the degree correlation of neighboring nodes, at the final state \cite{Newman2003a}. The initial state has negative assortativity because the hub-zealots are mostly connected to lower degree nodes at $t=0$. The secondary maximum coincides with the peak of the coefficient of variation of the assortativity (not shown) and low values of its mean. Thus, the network is \emph{on average} uncorrelated in degree, while for the same $\phi$, some simulations exhibit negative degree correlations while others show positive ones. The negative degree correlations imply that hubs, which are mostly zealots due to their excess degree $k_x$, are mainly connected to low degree non-zealot nodes. Because randomly chosen nodes mimic the opinion of a randomly chosen neighbor, the opinion of high degree nodes gets imitated more often, resulting in a faster spread of the zealot opinion and a shorter $t_c$. In turn, positive degree correlations result in a longer $t_c$, since in this situation, hubs are frequently connected with other hubs and, thus, their opinion $g_z$ reaches other hubs (mostly being zealots with the same opinion themselves) with elevated probability in a short amount of time, but takes much longer to affect the rest of the graph. Consequently, the secondary peak of $V_t$ indicates the rewiring probability that is necessary to compensate for the effect of the introduced degree heterogeneity. Accordingly, the coexistence of both types of degree correlations triggers a large $\sigma_{t_c}$ and therefore a large $V_t$. However, this effect has minor relevance for the ZOS because the secondary maximum of $V_t$ is located below the fragmentation threshold.

\subsection{Cluster size distributions}

\begin{figure}
  \centering
  \includegraphics[width=1.0\linewidth]{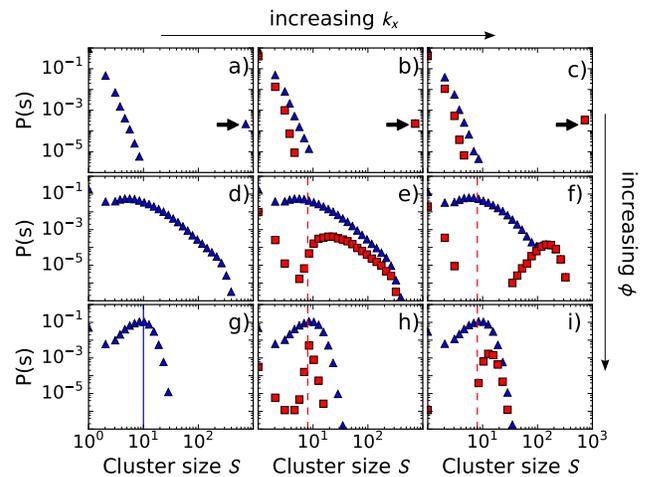}
  \caption{(Color online) Frequency distributions of cluster size $P(s)$ at $t=t_c$ for all clusters 
  (blue triangles) and for the fraction of convinced clusters (red squares) computed from $n=10^4$ simulation runs on graphs with $N=800$ and $\gamma_0=10$.
  The distributions are computed for no zealots $n_z^0=0$ (a,d,g),
  for zealot density $n_z^0=0.01$ and no excess degree $k_x=0$ (b,e,h), and for 
  zealot density $n_z^0=0.01$ and excess degree $k_x=20$ (c,f,i).
  The rows represent different values of the rewiring probability $\phi$: 
  $\phi = 0.04\ll\phi_c$ (a,b,c), $\phi=\phi_c(\overline{k})$ (d,e,f) and $\phi=0.96\gg\phi_c$ (g,h,i). Blue solid and red dashed vertical lines indicate $\gamma_0=10$ and initial number of zealots $N n_z^0$, respectively. Note the black arrows indicating the giant component.}	
  \label{fig:2}
\end{figure}

The most straightforward approach to maximize the ZOS efficiency $n_{z}(t_c)$ is to dominate the largest connected component in the final state. In the following, we will refer to connected communities with homogeneous node state (opinion) as clusters. In Fig.~\ref{fig:2}, the resulting frequency distribution $P(s)$ of cluster sizes $s$ (i.e., the number of nodes in a cluster) is shown for rewiring probabilities below (a,b,c), close to (d,e,f) and above the fragmentation threshold (g,h,i) of the three cases presented in Fig.~\ref{fig:1}(a,b,c).

The distributions without zealots are characterized by a giant component below
the critical point $\phi_c$ (Fig.~\ref{fig:2}(a)), by a power law behavior close to the transition (Fig.~\ref{fig:2}(d)) and by the absence of a giant component above the
fragmentation threshold (Fig.~\ref{fig:2}(g)), where the cluster size $s$ is distributed around $\gamma$, indicated by a vertical line in Fig.~\ref{fig:2}(g). Since no zealots and therefore no opinion $g_z$ are present the ZOS efficiency is always zero. 

Below the fragmentation transition, zealots dominate the giant component in both cases with and without excess degree (Fig.~\ref{fig:2}(b,c)) and on average spread their opinion to a fraction of $n_{z}\approx0.98 N$ nodes. Close to $\phi_c$, the distribution of clusters having the zealots' opinion $g_z$ (red squares) without excess degree (Fig.~\ref{fig:2}(e)) is similar to the total distribution (blue triangles) with the difference that convinced clusters have a size of at least $Nn_z^0$, indicated by a red dashed line. The excess degree (Fig.~\ref{fig:2}(f)) causes a peak at larger cluster sizes. The size distribution of convinced clusters starts at a size significantly larger than $n_z^0$ and coincides at larger cluster sizes with the total distribution, which shows that the corresponding maximum consists solely of convinced clusters. At the fragmentation transition the cluster distribution is expected to obey a power law (Fig.~\ref{fig:2}(d,e)), which is disturbed by the peak if zealots posses an excess degree (Fig.~\ref{fig:2}(f)). Consequently, it can be assumed that the excess degree is splitting the system into $g_z$-dominated subgraphs which have a larger mean degree than subgraphs that are not influenced by the zealots' opinion and excess degree. Hence, the $g_z$-dominated subgraphs effectively lie below or close to the local fragmentation transition and, therefore, tend to form larger clusters (Fig.~\ref{fig:2}(f), red squares). These subgraphs give rise to a five times larger ZOS efficiency as compared to the case of zealots without excess degree (Fig.~\ref{fig:2}(e)).

 The $g_z$-dominated subgraphs are also present above the fragmentation threshold, which is indicated by the fact that the previously discussed maximum of the cluster size distribution does not vanish suddenly but is gradually shifted towards smaller cluster sizes (Fig.~\ref{fig:2}(i)). In the case without excess degree (Fig.~\ref{fig:2}(h)), far above the transition point the size distribution of convinced clusters exhibits a maximum at $n_z^0$ (red line), while in the case with excess degree (Fig.~\ref{fig:2}(i)) this maximum is shifted towards larger cluster sizes, resulting in almost twice as high ZOS efficiency as compared to the case with $k_x=0$.

In summary, the cluster size distributions reveal that ZOS close to and above the
fragmentation transition is more efficient if hub-zealots are present due to a formation of convinced subgraphs with a larger local $\overline{k}$. Below the transition point, the mere existence of zealots is sufficient to reach a maximum ZOS efficiency.

\begin{figure}
  \hspace*{-0.4cm}
  \includegraphics[width=1.1\linewidth]{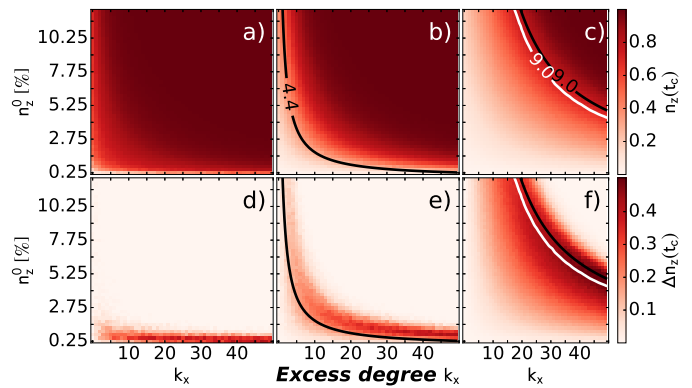}
  \caption{(Color online) (a,b,c) Fraction of convinced nodes at the final state $n_z(t_c)$ and (d,e,f) the former minus $n_{z\overline{k}}(t_c)$ in which the $k_x N n_z^0$ edges are distributed homogeneously instead to the $N n_z^0$ zealots (no zealot-hubs) resulting in $\Delta n_z(t_c)=n_z(t_c)-n_{z\overline{k}}(t_c)$, both as a function of excess degree $k_x$ and zealot density $n_z^0$. 
  The rewiring probability is increased from (a,d) $\phi=0.4$ via (b,e) $\phi=0.5$ to (c,f) $\phi=0.75$.   The black (b,c,e,f) and white (c,f) lines indicate contours where $\overline{k}=\overline{k}_c$ and $\overline{k}_z=\overline{k}_c$, respectively, highlighting the fragmentation transition, with $\overline{k}_z$ being the mean degree of $g_z$ nodes.
  All results have been obtained from $200$ simulation runs on networks with $N=800$ nodes and a mean 
  degree without zealots of $\overline{k}_0=4$.}
  \label{fig:3}
\end{figure}

\subsection{Systematic parameter study}

So far, the only parameter settings considered have been no zealots 
and zealots with initial density $n_z^0=0.01$ with and without excess degree $k_x=20$.
We now study a much wider set of parameters regarding the 
ZOS at final state $n_z(t_c)$ (Fig.~\ref{fig:3}).
The horizontal and vertical axes in Fig.~\ref{fig:3} represent the excess degree $k_x$ and density $n_z^0$ of zealots. 
The parameters $\overline{k}_0=4$ and $N=800$ are kept constant. Three different rewiring probabilities
are considered:
$\phi=0.4<\phi_c(\overline{k}_0)$ (Fig.~\ref{fig:3}(a)), $\phi=0.5> \phi_c(\overline{k}_0)$
(Fig.~\ref{fig:3}(b)) and $\phi=0.75 > \phi_c(\overline{k}_0)$ (Fig.~\ref{fig:3}(c)). Note that 
the fragmentation threshold depends on both $k_x$ and $\phi$, and the critical rewiring probability $\phi_c(\overline{k}_0)=0.45$ 
can therefore only be considered as a reference point for $k_x=0$. Thus, for a fixed $\phi$
there exists a critical mean degree $\overline{k}_c$. If this mean degree is exceeded, the system is below its fragmentation threshold, and, hence, in a regime where ZOS profits from the presence of a giant component. 

Below the fragmentation transition (Fig.~\ref{fig:3}(a)), we find that an increase in either the excess degree or the number of zealots strongly increases the ZOS efficiency, which quickly maximizes because a giant component is easily dominated. Sufficiently far below the transition point (e.g., for $\phi=0.1$, not shown), every individual configuration results in maximum ZOS.

Just above the transition point (Fig.~\ref{fig:3}(b)), the presence of additional zealots increases the ZOS efficiency as expected. However, ZOS efficiency quickly saturates to a maximum if together with an elevated excess degree, the critical degree $\overline{k}_c=4.4$ is reached where the giant component fragments.

Finally, far above $\phi_c(\overline{k}_0)$ (Fig.~\ref{fig:3}(c)), the fragmentation transition is reached at $\overline{k}_c=9$ (black contour line)
with, of course, larger amounts of zealots and excess degrees. 
It is remarkable that already below the corresponding transition point,
ZOS efficiency increases to multiples of $n_z^0$. This effect can be explained by the convinced clusters 
that have larger mean degree than unconvinced ones (compare Fig.~\ref{fig:2}(f)). 
Note that the mean degree of $g_z$ nodes $\overline{k}_z$
at the final state crosses $\overline{k}_c$ already at a smaller zealot density and excess degree than the total mean degree (the case $\overline{k}_z=\overline{k}_c$ is highlighted by a white contour line in Fig.~\ref{fig:3}(c)). 
Those subgraphs with larger mean degree are a key finding of this study and can be interpreted
as a community in which intense discourses were triggered by the excess degrees of the 
zealots and their opinion. Note that the critical mean degree $\overline{k}_c$ is estimated by reproducing Fig.~\ref{fig:1}(a) with a different $\overline{k}_0$ until the peak of $V_t$ coincides with the value of $\phi$ used in Fig.~\ref{fig:3}(b,c), respectively.    

In order to properly interpret the results discussed above, it is important to understand to which extend the observed emergence of densely connected subgraphs responsible for ZOS efficiency originates from the presence of zeolots with distinct excess degree as opposed to mere effects of an elevated mean degree $\overline{k}$ of the whole network. For this purpose, Fig.~\ref{fig:3}(d,e,f)) presents the difference $\Delta n_z(t_c) = n_z(t_c) - n_{z\overline{k}}(t_c)$ between the ZOS efficiency $n_z(t_c)$ of our AVM variant as discussed above and the ZOS efficiency $n_{z\overline{k}}(t_c)$ that would arise if the additional $k_x N n_z^0$ links were distributed homogeneously among all nodes of the networks instead of assigning them exclusively to the zealots. This difference is always positive and largest in the non-fragmented phase close to the transition point (Fig.~\ref{fig:3}(e))). However, at large rewiring probabilities ($\phi=0.75$, Fig.~\ref{fig:3}(f)), the parameter range for which marked differences between both settings are found extends over large parts of the parameter subspace corresponding to the non-fragmented phase. These findings demonstrate that the emergence of subgraphs with increased mean degree is mainly caused by the presence of hub-zealots. 

We emphasize that qualitatively and quantitatively similar results are observed (not shown) for situations in which the excess degree is distributed among randomly selected nodes of the network (random hubs) or among randomly chosen non-zealots only (non-zealot hubs). In fact, the results for the random hubs are quantitatively extremely similar to the case with homogeneously distributed edges (Fig.~\ref{fig:3}(d,e,f)). Thus, we conclude that it does not matter much if the mean degree is increased homogeneously or heterogeneously as long as the increase does not favor nodes of a specific opinion. In turn, marked differences emerge if the excess degree is fixed to nodes of a single opinion (in our case, the zealots).

\section{Macroscopic approximation \label{sec:ana}}

For a model similar to that studied in the present work~\cite{Wiedermann2015}, it was recently shown that a mean-field approximation can be performed to derive analytical results by considering solely pairwise interactions and 
under the assumptions that the network is large and fully connected at initiation. 
In the latter work, only two distinct node states (aka opinions) have been present. This assumption is adopted in the following by treating all opinions different from $g_z$ as equivalent, i.e., as the opinion of the others $g_o$. Following~\cite{Wiedermann2015}, this simplification reduces the problem to three coupled differential equations for the time evolution of three macroscopic properties of the model: the fraction of nodes $n_z$ holding opinion $g_z$ and the average numbers of links per node $m_{zz}$ ($m_{oo}$) among nodes holding opinion $g_z$ ($g_o$):

\hspace*{-1.2cm}
\begin{align}
\frac{dn_z}{dt} &= (1-\phi)\left[n_o P_{o}^z - (n_z-n_z^0) P_{z}^o\right], \label{eq:nz}\\
\frac{dm_{zz}}{dt} &= \phi n_z P_{z}^o + (1-\phi) 
\left[P_{o}^z m_{zo} - \frac{n_z-n_z^0}{n_z} 2 P_{z}^o m_{zz} \right], \label{eq:mz} \\
\frac{dm_{oo}}{dt} &= \phi n_o P_{o}^z + (1-\phi) 
\left[\frac{n_z-n_z^0}{n_z} P_{z}^o m_{zo} -  2 P_{o}^z m_{oo} \right]. \label{eq:mo}
\end{align}
Here, $m_{zo}=M_{zo}/N$ where $M_{zo}$ is the number of ``active'' links between nodes of distinct opinions $g_z$ and $g_o$, respectively. Note that the excess degree $k_x$ enters the above equation via the fraction of initially active links $m_{zo}(t=0)=n_z^0(\overline{k}_0+k_x)$.
$P_z^o$ is the probability of a $g_z$ node to interact with a $g_o$ node and is given by the heterogeneous mean-field approximation~\cite{Wiedermann2015} as 
\begin{equation}
P_{z}^o = \frac{k_z^o}{k_z} = \frac{m_{zo}}{2m_{zz}+m_{zo}}.
\end{equation}
Here, $k_z$ is the mean degree of $g_z$ nodes and $k_z^o$ is the mean number of links from a $g_z$ node to $g_o$ nodes. $P_o^z$ follows analogously by exchanging the indices.

Equation~(\ref{eq:nz}) implies an increase of $n_z$ by $g_o$ nodes being convinced by $g_z$ nodes and a decrease through non-zealot $g_z$ nodes being convinced by $g_o$ nodes. 
The first term in Eq.~(\ref{eq:mz}) describes an increase of $m_{zz}$ by $g_z$ nodes cutting their link to $g_o$ nodes. The second term consists on the one hand of
an increase due to $g_o$ nodes becoming convinced, transforming the mean per-node number of links between different opinions from $m_{zo}/n_o$ to $m_{zz}$, and on the other hand a decrease by non-zealot $g_z$ nodes changing their opinion and, thus, transforming the mean number of $m_{zz}$ links of one $g_z$ node from $2 m_{zz}/n_z$ to $m_{zo}$ (analogously for Eq.~\eqref{eq:mo}).
Also note that Eqs.~(\ref{eq:nz})--(\ref{eq:mo}) are closed, i.e., $m_{zz}+m_{oo}+m_{zo}=\overline{k}/2$ and $n_z+n_o=1$.
The main difference to the model in~\cite{Wiedermann2015} is the presence of zealots included by reducing the fraction $n_z$ in the convincing process by $n_z^0$.
 
For the model considered here, five fixed points can be identified as unstable or outside the regime of interest 
($0\leq n_z^0\leq n_z\leq 1$, $0\leq m_{zz}+m_{oo}\leq \overline{k}/2$). 
A two-dimensional manifold, which represents the consensus/final state, also satisfies
the stationarity criterion:
\begin{equation}
m_{zz}^\star = \frac{\overline{k}}{2}-m_{oo}\ \ \ (m_{zo}=0).
\label{eq:mani}
\end{equation}
Note that this manifold extends over all values of $n_z$ and $m_{oo}$.
Its linear stability properties are determined by the eigenvalues of the Jacobian at $m_{zz}^\star$, which are $0$, $0$ and 
\begin{equation}
\begin{split}
 &f(n_z, m_{oo},n_z^0,\overline{k},\phi) =\\
 &\qquad \frac{\phi n_z}{2m_{oo} -\overline{k}} +
 \frac{2 m_{oo} (n_z^0-2 n_z) (\phi-1)+(n_z-1) n_z \phi}{2 m_{oo} n_z}.
\end{split}
\end{equation}
Since two eigenvalues are zero, the manifold cannot be asymptotically stable but (un-)stable if $f(n_z, m_{oo},n_z^0,\overline{k},\phi)<0$ $(>0)$.
In the following, it is assumed that the links are, regardless of the initial conditions, homogeneously distributed at $t_c$, i.e., $m_{oo}^\star=\overline{k}(1-n_z)/2$, which simplifies the nonzero eigenvalue to 
\begin{equation}
 f(n_z, m_{oo}^\star,n_z^0,\overline{k},\phi)=2-\frac{n_z^0 (1-\phi)}{n_z}-\frac{2 (1+\overline{k}) \phi}{\overline{k}}.
 \label{eq:manevs}
\end{equation}
\noindent
The latter changes its sign at
\begin{equation}
 \tilde{\phi}_c(n_z,n_z^0,\overline{k})= 1- \frac{2 n_z}{2 (1+\overline{k}) n_z-\overline{k} n_z^0}.
 \label{eq:phic1}
\end{equation}
Thus, the transition depends on $\overline{k}$ such that $\lim_{\overline{k}\to 0} \tilde{\phi}_c=0$
and $\lim_{\overline{k}\to \infty} \tilde{\phi}_c=1$. 
The macroscopic approximation is strictly valid in the case $N\to\infty$; otherwise, finite-size effects may lead to deviations of the system from the analytically calculated behavior.

The fragmentation threshold is approximated by $\tilde{\phi}_c(n_z=1)$, because 
at the transition weakly connected clusters split up regardless of their size
and $\tilde{\phi}_c$ increases with $n_z$. This results in
\begin{equation}
 \tilde{\phi}_c(1,n_z^0,\overline{k})= 1- \frac{1}{1+\overline{k}(1- n_z^0\slash 2)}.
 \label{eq:phic2}
\end{equation}
Note that for $n_z^0\ll1$ the transition can be considered as independent of $n_z^0$,
which is in perfect agreement with the numerical findings in Fig.~\ref{fig:1} with $n_z^0=0.01\ll1$, where the transition is not shifted by changing $n_z^0$ (Fig.~\ref{fig:1}(b)) but by a change in $k_x$ (Fig.~\ref{fig:1}(c)), since $\overline{k}=\overline{k}_0+2n_z^0k_x$. 
\begin{figure}
  \hspace*{-0.7cm}
    \includegraphics[width=1.1\linewidth]{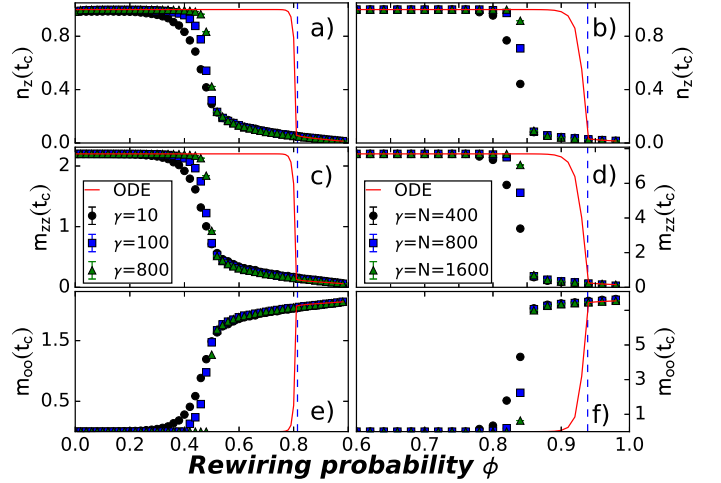}
  \caption{(Color online) Results of microscopic ensemble simulations (symbols) and macroscopic approximation (solid lines) using Eqs.~(\ref{eq:nz})--(\ref{eq:mo}).
  Mean values have been estimated from $200$ simulation runs. (a) ZOS efficiency $n_z(t_c)$, (c) density of links between convinced nodes $m_{zz}(t_c)$, (e) density of links between unconvinced nodes $m_{oo}(t_c)$ at convergence time $t_c$ with $N=800$ and varying $\gamma$ as indicated in (e). (b,d,f) Same as for (a,c,e) but with varying $N=\gamma_0$ and therefore only one opinion different from $g_z$. The blue dashed vertical line marks the approximate phase transition $\tilde{\phi}_c$ (Eq.~(\ref{eq:phic2})). For all runs, $n_z^0=0.01$ and $k_x=20$ have been kept fixed. In (a,c,e), $\overline{k}_0=4$, while in (b,d,f), $\overline{k}_0=15$. Note the different scales along the horizontal axes.}
  \label{fig:4}
\end{figure}

We compare the mean-field model and the microscopic model by computing the final state values of $n_z$, $m_{zz}$ and $m_{oo}$ for Eqs.~(\ref{eq:nz})--(\ref{eq:mo}) by forward integration and by taking an average over the outcomes of $200$ simulations of the microscopic model (Fig.~\ref{fig:4}). Both, the excess degree with $k_x=20$ and the zealot density with $n_z^0=0.01$, are kept constant across all simulations. A low mean degree $\overline{k}_0=4$ is considered in  Fig.~\ref{fig:4}(a,c,e). Simulations with this setting show a fragmentation transition at $\phi_c\approx0.5$, which disagrees with the analytical approximation of Eq.~(\ref{eq:phic2}), $\tilde{\phi}_c\approx 0.814$, marked by a dashed vertical line, by a relative error of $\delta\phi_c=0.38$. This large discrepancy is to be expected since 
a large mean degree is necessary for the mean-field approximation to reach a good agreement with the full numerical results of microscopic simulations~\cite{Gleeson2012}.

In Fig.~\ref{fig:4}(b,d,f) the mean degree is increased to $\overline{k}_0=15$. The numerical results indicate $\phi_c\approx0.85$, while the analytics give 
$\tilde{\phi}_c\approx0.94$, which reduces the discrepancy to a relative error of $\delta\phi_c=0.096$. The remaining disagreement can very likely be further reduced by including higher-order terms in the AVM~\cite{Kimura2008}, which is a promising task for future research. Note that far above and below the transition point, both modeling approaches agree well with each other.

Additionally, in Fig.~\ref{fig:4}(a,c,e) the validity of the simplification of considering all opinions other than $g_z$ the same as $g_o$ is checked by decreasing $\gamma$, which increases the diversity of opinions. Below but close to the phase transition point, simulations with a higher diversity show a reduced ZOS efficiency due to the higher probability of small groups to cluster and cut their links to the giant component, cf.~the black circles in Fig.~\ref{fig:4}(a). Above and sufficiently below the fragmentation transition, the obtained results are robust for different $\gamma$.

In Fig.~\ref{fig:4}(b,d,f) we check for finite size effects by simulating at $N=400$, $800$ and $1,600$ while keeping the diversity of opinions at its minimum ($\gamma=N$). The finite size smoothens the transition from the giant component to the fragmented phase, as can be seen in Fig.~\ref{fig:4}(b). Here, simulations with the largest number of nodes (green triangles)
switch more suddenly to the fragmented phase (compare black circles and green triangles) than such for smaller networks.

We note that the excess degree is responsible for the observed slow decrease of $n_z(t_c)$ after the fragmentation transition until it reaches $n_z^0$ at $\phi=1$ (Fig.~\ref{fig:4}(a,b)). If the $k_x N n_z^0$ additional links are instead distributed homogeneously among all nodes, resulting in an elevated degree without distinct zealot hubs, the ZOS efficiency is markedly reduced already at the transition point (not shown), which can also be seen in our microscopic simulations for the ZOS efficiency difference between the two cases in Fig.~(\ref{fig:3}(e,f)) close to the transition point.

%
%

\section{Alternative Zealot Update Schemes \label{sec:altup}}

In the previous sections, we have treated zealots as different from normal agents exclusively in terms of their additional property to be stubborn. Doing so, we have been in line with vast parts of the existing literature.  However, considering that real world zealots can be expected to aim to maximize the spread of their respective opinion, it appears unintuitive that they should follow almost the same update rules as non-zealots. Therefore, in the following we investigate two alternative update schemes, which should potentially increase the ZOS efficiency. For brevity, we leave any in-depth analysis of the resulting dynamics along the lines of the previous sections as a subject of future research.  

As a first possible scenario, we investigate the case of \textit{passive} zealots that do not rewire at all and, thus, do not cut links to $g_o$-nodes. This procedure then enhances the probability to convince the latter. In this case, the second step in the update cycle as presented in Sec.~\ref{sec:model} is modified such that

\vspace{0.2cm}
\begin{minipage}{0.95\columnwidth}
\noindent
Step 2: If $i\in \mathcal{S}_z$ do nothing, otherwise follow steps 2(a) and (b).
\end{minipage}
\vspace{0.2cm}

Passive zealots can be interpreted as agents who are forced to a trade-off between the large number of connections kept to other agents and their ability to rewire to new agents. They benefit from their large degree but in turn cannot manage to establish new links.

As a second case, we study \textit{heterophilic} zealots that rewire exclusively to $g_o$ nodes. This restriction is expected to further enhance the efficiency of convincing the latter. Accordingly, the second step in the update scheme of Sec.~\ref{sec:model} is then modified as follows:

\vspace{0.2cm}
\begin{minipage}{0.95\columnwidth}
\noindent
Step 2(a): With probability $\phi$, if $i\in \mathcal{S}_z$ and $|\mathcal{S}_i|\neq 0$, delete the link to $j$ and rewire to a randomly selected node of the set $\mathcal{S}_{g_o}\setminus\mathcal{N}_i$. If $i\notin \mathcal{S}_z$ and $|\mathcal{S}_i|\neq 0$, rewire to a node of the set $\mathcal{S}_i$. Otherwise, do nothing.
\end{minipage}
\vspace{0.2cm}

Former studies on AVM variants with heterophily found that the fragmented phase vanishes \cite{Kimura2008}. However, in contrast to these previous works which introduced heterophily to all nodes, we here only declare zealots as being heterophilic, whereas normal nodes remain homophilic. Thus, the fragmented phase can still be reached if there exist no links to any $g_z$-node, i.e., $m_{zz}=m_{zo}=0$.

\begin{figure}
  \hspace*{-0.5cm}
    \includegraphics[width=1.1\linewidth]{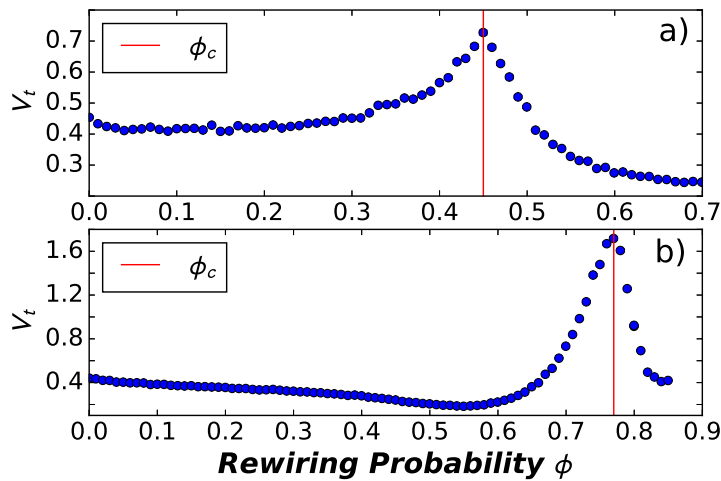}
  \caption{(Color online) Coefficient of variation of convergence time $V_t$, 
    indicating a phase transition at $\phi=\phi_c$ where
    the giant component vanishes. \textit{Passive} (a) and \textit{heterophilic} (b) zealots are  
    compared. 
    The red line in (a) marks $\phi_c=0.45$ and in (b) $\phi_c=0.77$. Distributions are computed from $n=10^4$ runs on graphs with $N=800$ nodes, a zealot density of
    $n_z^0=0.01$ with no excess degree $k_x=0$ and a mean 
    degree of $\overline{k}_0=4$.}
  \label{fig:5}
\end{figure}

In Fig.~\ref{fig:5}, the coefficient of variation of the convergence time, $V_t$, is analyzed for both types of zealots for the case without excess degree only, since we focus here on the effect of different zealot update schemes rather than on the effect of hub zealots (cf.\ Fig.~\ref{fig:1}(b)). Interestingly, $\phi_c\approx 0.45$ remains valid for the case of \textit{passive} zealots (Fig.~\ref{fig:5}(a)).
Furthermore, the difference $\Delta n^0_z(t_c)$ between the ZOS efficiency in the case of \textit{passive} and \textit{normal} zealots, as shown in Fig.~\ref{fig:3}(d,e,f), does not show any differences between the two rewiring schemes over a wide parameter range (not shown).

A possible explanation for the aforementioned finding is that, as expected, the unrewired links to $g_o$ nodes increase the probability to interact with $g_o$ because of an increase in $m_{zo}$, while keeping $m_{oo}$ constant increases $P_o^z$ and therefore the first term in Eq. ~(\ref{eq:nz}), resulting in larger values of $n_z$. However, $m_{zz}$ is reduced in the absence of homophilic rewiring of the zealots. Assuming that $m_{zz}$ is decreased by the amount of links which increases $m_{zo}$ results in an increase of $P_z^o$ which reduces $n_z$, as can be seen from the second term of Eq.~(\ref{eq:nz}). In fact, $P_z^o$ also increases if $m_{zz}$ stays constant. Thus, these two processes compete against each other, which could explain the observed effect. However, in this perspective, we have neglected (besides other aspects) that the considered missing $m_{zz}$ links connect to zealots. Thus, non-zealot $g_z$ nodes are more likely to be convinced and stay $g_o$ nodes than in the setting involving \textit{normal} zealots.

In contrast to the aforementioned behavior, the fragmentation transition for \textit{heterophilic} zealots (Fig.~\ref{fig:5}(b)) shifts strongly to $\phi_c\approx 0.77$. 
This shift causes an increase of ZOS efficiency, since a giant component is present over a larger parameter range in $\phi$. This increased stability of the giant component is caused by the mixed behavior of the $g_z$ nodes: while the non-zealots exhibit homophilic rewiring and therefore stay connected to the zealot $g_z$ nodes, the zealots themselves reach out for other $g_o$ nodes, which could already be parts of an isolated component consisting only of like-minded nodes. These isolated components were not reachable for \textit{normal} zealots, whereas \textit{heterophilic} zealots can ``invade'' these groups of nodes and convert parts of them or even the whole groups. Note that non-zealot $g_z$ nodes provide the zealots with links via their homophilic rewiring.

In summary, the two alternative zealot rewiring schemes discussed in this section illustrate that there is a large class of update schemes, which perform equally well or even better than the one discussed in the previous sections in terms of ZOS efficiency. Especially the strong effect of \textit{heterophilic} zealots and how non-zealot $g_z$ nodes provide links to them to enable the zealots to connect to not like-minded nodes demonstrate that there can exist interesting feedback effects between two coexisting rewiring schemes, which seems a promising field of future research.

\section{Conclusions\label{sec:discuss}}

In this paper, we have introduced zealots with an increased mean degree, the excess degree, into the adaptive voter model (AVM) and investigated how their fixed, uniform and new opinion spreads over a social network until a full or fragmented final state is reached. The efficiency of zealot opinion spreading (ZOS) has been quantified by the fraction of nodes holding the zealots' opinion at the asymptotic state.

After reproducing the results of a previous study~\cite{Holme2006} by means of numerical simulation, a detailed comparison of the resulting cluster size distributions below, at and far above the fragmentation transition revealed the existence of zealot-dominated subgraphs if the introduced zealots exhibit an excess degree. These subgraphs are characterized by an elevated mean degree and the presence of opinions enforced by the hub-zealots.

By investigating a wide range of initial zealot densities and excess degrees,
three regimes were identified in which different effects allow a maximum or an increased ZOS efficiency.
Below the fragmentation transition, maximum ZOS is easily achieved by 
introducing solely a few zealots without excess degree due to the emergence of a giant component, which allows for ZOS across the whole network.
Shortly above the fragmentation transition, an increase of zealot density is insufficient to increase ZOS efficiency.
In addition, an excess degree needs to be introduced and raised to a certain level to allow for the system returning to the non-fragmented phase in which ZOS efficiency is quickly maximized. Thereby, either the mean degree of a large amount of zealots is slightly increased by the excess degree, or a small amount of zealots is declared as hubs. 
Far above the phase transition, only a large density of hub-zealots is able to push the system to the non-fragmented phase.
However, the formation of zealot-dominated subgraphs with increased mean degree plays a crucial role
and allows for increased ZOS efficiency already far below the phase transition point.
Since these subgraphs emerge in the fragmented phase, they only include a specific fraction of nodes, determined by the cluster size distribution.

We have macroscopically approximated the model by considering pairwise interactions. An analytical approximation of the
phase transition point was found, which was validated by forward integration.
However, the theoretically approximated critical point is much larger than suggested by the numerical simulation of the microscopic model.
This discrepancy was reduced by considering systems with larger mean degree. Previous studies~\cite{Kimura2008} suggest the potential for further improvement if the approximation is not only based on pairwise interactions. However, far below and above the phase transition point,
analytics and numerics agree well with each other.

Finally, we have studied the effect of two alternative update schemes for the zealots. Here, it was shown that \textit{passive} zealots, which do not rewire, perform as good as their normal counterparts in terms of ZOS efficiency. In turn, \textit{heterophilic} zealots, which rewire to not like-minded nodes only, shift the fragmentation transition strongly to larger rewiring probabilities and therefore have a much larger ZOS efficiency. These update schemes provide an interesting starting point for future research, which should investigate the effects of a coexistence of different update rules in the same model.

The finding of zealot-dominated subgraphs with a larger mean degree than in the rest of the graph in the fragmented phase allows drawing the following conclusion: Large communities (sub-networks) with agents engaging in active discourse (larger mean degree than in other communities) are likely to be targeted by the interests and the resources of an already convinced group. In our model, an active discussion can imply that there is an attempt to control the system. 
In order to avoid being controlled by an external opinion, each node should keep links to nodes from other subgraphs.

How to maximize or minimize opinion spreading by zealots is a relevant question in the context of the AVM as well as related models of opinion dynamics.
Similar studies already focused on the static voter model~\cite{Wu2004, Yildiz2013} or on general 
dynamical systems~\cite{Liu2011a, Cowan2012, Nacher2012, Sun2013, Yuan2013}
on static networks. In the latter research, it was of crucial interest if there exist specific nodes which have topologically favorable or unfavorable positions to spread their opinion, or if there exists a minimum set of nodes necessary to 
spread an opinion across the whole network. Along these lines, it is of interest to quantify the effect of network topology on ZOS efficiency in the AVM in future research. An increase of the rewiring probability changes the topology faster and increases (below the fragmentation transition) the convergence time. Consequently, the larger the rewiring probability $\phi$, the stronger the initial topology is modified. Hence, topological effects are only expected to play a role at low rewiring probabilities. However, as shown by our study for random graphs, in this regime opinion spreading
is already maximized by randomly positioning a small number of zealots without excess degree.
For networks with a more complex topology than a random graph, the corresponding effect might be quantitatively or even qualitatively different.
Above the transition point, especially the topological effects on cluster formation with increased mean degree are of interest.

More generally, this study has also presented a strong motivation for further investigating the controllability of the AVM as an example for a nonlinear dynamical system on a dynamic network. This is because the ZOS efficiency, on which we have focused our interest, does not represent exactly the controllable subset of the zealots, but might still be closely related to this concept. If the ZOS efficiency would represent the subset of nodes controllable by the zealots, we could drive this subset from any initial state to any desired state within a time $t_c$ by an appropriate input signal. However, the ZOS efficiency only shows that we drove the subset to a specific, not an arbitrary, state.

This study has been based on the approach of intervening in the opinion adoption process in the AVM. Clearly, a complementary approach would be to interfere with the rewiring process, which could imply to declare specific links as unbreakable or harder to break. Also, the creation of links between zealots by rewiring could be
excluded, assuming that campaigners or lobbyists have no interest in clustering among themselves. First, it would
be easier to identify them as zealots and second, they would waste their linkage resource needed for influencing other nodes.

Another way of intervening in the opinion formation process is to change the update rules for the zealots. In this case, we have already studied two model variants with \textit{passive} and \textit{heterophilic} zealots, whereby the latter results in a large increase of ZOS efficiency. A former study investigated how agents maximize their power (represented by a score function increasing with centrality and decreasing with degree, the \textit{diplomats dilemma}) by following specific link creation and deletion strategies, assuming knowledge up to the second neighborhood \cite{Holme2009}. The most common strategy to increase the corresponding performance was to delete the link to the direct neighbor with largest centrality and add an edge to the neighbor in the second neighborhood with largest centrality. This could be one further update rule, next to many others. In this context, an interesting question would be to identify specific update schemes which are robust across different network types.

In summary, we emphasize that the combination of intervening in opinion adoption and rewiring processes, different zealot update schemes, considerations regarding the role of complex network topology and generalizing those approaches to more realistic models of social network dynamics~(e.g., \cite{Schleussner2016a, Barfuss2016}) are promising fields of future research.

%
%
%
%

\appendix

\begin{acknowledgments}
P.P.K., M.W. and R.V.D. have been financially supported by the German Federal Ministry of Education and Research (BMBF) via the Young Investigators Group CoSy-CC$^2$ (grant no. 01LN1306A). J.F.D. thanks the Stordalen Foundation (via the Planetary Boundary Research Network PB.net) and the Earth League's EarthDoc program for financial support. M.W and J.F.D thank the Leibniz Association (project DOMINOES) for financial support. The authors gratefully acknowledge the European Regional Development Fund (ERDF), the German Federal Ministry of Education and Research and the Land Brandenburg for supporting this project by providing resources on the high performance computer system at the Potsdam Institute for Climate Impact Research. The presented research was conducted within the scope of the COPAN flagship project on co-evolutionary pathways at the Potsdam Institute for Climate Impact Research (\url{http://www.pik-potsdam.de/copan}) and the International Research Training Group IRTG 1740/TRP 2014/50151-0, jointly funded by the German Research Foundation (DFG, Deutsche Forschungsgemeinschaft) and the S\~{a}o Paulo Research Foundation (FAPESP, Funda\c{c}\~{a}o de Amparo \`a Pesquisa do Estado de S\~{a}o Paulo). 
\end{acknowledgments}

\bibliography{library2}

\end{document}